\documentclass[12pt,floats]{iopart}
\usepackage{epsfig}

\def\no{\nonumber}

\def\t{\tau}

\def\om{\omega}
\def\be{\begin{equation}}
\def\bea{\begin{eqnarray}}
\def\eea{\end{eqnarray}}
\def\ee{\end{equation}}
\def\bi{\begin{itemize}}
\def\ei{\end{itemize}}

\def\d{\delta}

\def\bn{\begin{enumerate}}
\def\en{\end{enumerate}}

\begin{document}

\title[Validating delta-filters for resonant bar....]
{Validating delta-filters for resonant bar detectors of improved
bandwidth foreseeing the future coincidence with interferometers}
\author{Sabrina D'Antonio \dag, Archana Pai \ddag, Pia Astone \ddag}

\address{\dag INFN Sezione Roma 2 Tor Vergata ITALIA}
\address{\ddag INFN Sezione Roma 1 --- P. le Aldo Moro, 2 00185 Roma ITALIA}

\ead{sabrina.dantonio@lnf.infn.it,archana.pai@roma1.infn.it, pia.astone@roma1.infn.it}

\begin{abstract}
The classical delta filters used in the current resonant bar
experiments for detecting GW bursts are viable when the bandwidth of resonant
bars is few Hz. In that case, the incoming GW burst is likely to be viewed
as an impulsive signal in a very narrow frequency window. After making 
improvements in the 
read-out with new transducers and high sensitivity dc-SQUID, 
the Explorer-Nautilus have improved the bandwidth ($\sim 20$ Hz) 
at the sensitivity level of $10^{-20}/\sqrt{Hz}$. Thus, 
it is necessary to reassess this assumption of
delta-like signals while building filters in the resonant bars as the
filtered output crucially depends on the shape of the waveform. This
is presented with an example of GW signals -- stellar
quasi-normal modes, by estimating the loss in SNR and the error in the
timing, when the GW signal is filtered with the delta filter as compared
to the optimal filter.
\end{abstract}

\pacs{04.80 Nn, 07.05 Kf, 97.80 -d}
\maketitle

\section{Introduction}
Till date, the burst data analysis in the narrow-band resonant detectors
like Explorer-Nautilus is based on the assumption that short gravitational
wave(GW) bursts (duration
of few millisecond to fraction of a second)
appear as delta-like signals in the detector bandwidth (BW). Such an assumption
was made mainly because the short GW bursts emit
waveforms of unknown shape and further due to detector's narrow band, one
could safely assume that the signal emits a flat spectrum in the detector BW.
However, recent improvements in resonant bar
detectors, mainly the read-out with new transducers and high coupling
dc-SQUID \cite{Pizzella:03}, have improved the BW ($\sim 20$ Hz)
at the sensitivity level of $10^{-20}/\sqrt{Hz}$, see Fig \ref{fig:exp}(a).
Thus, it is important to reassess the above assumption which we
demonstrate in this paper with an example of stellar quasi-normal modes (QNM).

Astrophysical inputs indicate \cite{Andersson:2004bi} that various
physical processes can excite stellar QNM during its evolutionary
phases, emitting GW in the BW of the resonant bars.  Such GW
may last for a fraction of a second to few seconds.  For example,
after the SN core collapse, during the cooling phase, the proto-NS
emits GW as a damped sinusoid signal evolving in frequency as well as
damping time and can also chirp in the resonant bars
\cite{Valeria:2003}\footnote{Possible candidates for coincidences
with interferometers like VIRGO.}. In this case, if the data is
filtered through a filter matched to a delta-like signal -- a delta
filter -- rather than a proper matched filter, the error in arrival
timing as well as loss in SNR can arise.  Here, as a preliminary
study, we discuss these issues for a simple case of stellar QNM
emitting GW $h(t)$ modeled as a damped sinusoid with a fixed
frequency $f_0=\om_0/2\pi$ and damping time $\t$, as given below
\be
\label{ht}
h(t) = h_0~\sin[\om_0 (t-t_0)]~e^{-(t-t_0)/\t}~\theta(t-t_0) \,.
\ee
Here, $\theta(t-t_0)$ implies $h(t) \neq 0$ for $t \ge t_0$ ($t_0$ is the time
of arrival) and zero otherwise. The damping time $\t$ depends on the underlying physical process.

To carry out the delta filter assessment, we assume the signal parameters
$f_0$ and $\t$ are known and compare the outputs of the delta filter
to that of the matched filter.

\section{System response}
The incoming $h(t)$ can excite the first 
longitudinal mode of the resonant bar. A low mass electrical transducer 
is attached to the bar to convert this displacement of the bar end face
into an electrical signal which is further
amplified with the SQUID amplifier. Such a bar-transducer 
system is modeled as a coupled harmonic oscillator with two resonant
modes \cite{Astone:1997dj}. The $h(t)$ provides an external force 
$f_x(t) = m_x L \ddot{h}(t)/2$ to the bar
[$m_x$: the reduced mass of the system $=M_{\rm bar}/2$,
$L$: the effective length $={\rm 4L_{\rm bar}}/\pi^2$].

The electrical output of the transducer is proportional to the
relative displacements of the transducer and the bar,
$u(t) = y(t) - x(t)$ from their equilibrium positions. In Fourier
domain \cite{Astone:1997dj}, the response $u(t)$ to the external force
$f_x(t)$ is obtained by 
\be
\label{uom}
U(j \om) = W_{ux}(j \om) F_x(j \om) \,, 
\ee
where $W_{ux}(j \om)$ is the system transfer function from the input force to
the output displacement as given defined in Eq.(1.8) of \cite{Astone:1997dj}.
The $F_x(j \om)$ [the FT of $f_x(t)$] for $h(t)$ defined in Eq.~(\ref{ht})
is
\be
\label{fx2}
\!\!\!\!\!\!
F_x(j \om) =  \frac{-m_x L \om^2}{2} H(j \om) = \frac{-m_x h_0 L \om^2}{2} \frac{\om_0 \t^2}{(1+j \om \t)^2 + \t^2 \om_0^2} \exp(-j \om t_0)
\ee
where $H(j \om)$ is the FT of $h(t)$. The $|F_x(j \om)|$ is a
Lorentzian which becomes narrower (broader), and in(de)creases in
height as $\t$ in(de)creases, see Fig. \ref{fig:exp}(a). Conveniently,
we choose $t_0 = 0$. The relative displacement of the transducer
$u(t)$ is in units of length.  The electrical signal at the output of
the SQUID amplifier is then given by
\be
\label{eq:vt}
v(t) = K u(t) \equiv M_v u_0(t) \quad \quad \quad \quad {\rm in~units~of~V}\,,
\ee
where $M_v \equiv K M_u$ and $M_u$ is the maximum value of $u(t)$ \footnote{$K$  is
equal to $\alpha A B$ of Ref. \cite{Astone:1997dj} which includes SQUID amplification and transducer constants.}.

The expected power spectral density (PSD) of the noise $n(t)$ at the output of the electrical chain is given by \cite{Astone:1997dj}
\be
S_t(\om) = S_n + \alpha^2 S_{fx} |W_{ux}(j \om)|^2 + \alpha^2 S_{fy} |W_{uy} (j \om)|^2 \,.
\ee
The $S_n$ is the broad-band noise contribution from the SQUID and the electrical chain
with a flat spectrum in the BW and the rest is the narrow-band noise contribution 
-- due to the
thermal noise of the two mechanical oscillators. The $S_{fx}$ and $S_{fy}$ is the 
total noise force spectra due to the Nyquist
and the back-action force. The $W_{ux}$ and $W_{uy}$ are the system transfer
functions as defined in \cite{Astone:1997dj} \footnote{Besides, the other
spurious unknown noise sources are treated while filtering the data.}.

\section{Matched filter}
\par
Signal detection problem involves computing a statistic -- 
a functional of the data $z(t)$  -- 
which when passed through the threshold allows to make the
decision of  
either presence [i.e. $z(t) = n(t) + v(t)$] or absence of 
signal [i.e. $z(t) = n(t)$] in the data.
\par
For known signal in Gaussian-stationary noise,
matched filtering is the optimal filtering.
The matched filter output is given by 
\bea
\label{eq:gt}
o(t) &\equiv& <z,q> = \frac{1}{2\pi} \int_{-\infty}^{\infty} Z(j \om) Q(j \om) \exp(j \om t) d \om \,.\eea
The matched filter transfer function for $v(t)$ is
%
$Q(j\om) = N_u U_0^*(j\om)/S_t(\om)$ with 
%
$U_0(j\om)$ as the FT of $u_0(t)$. The normalization 
$N_u = [\frac{1}{2\pi} \int_{-\infty}^{\infty} |U_0(j\om)|^2/S_t(\om) d \om]^{-1}$ is such that ${\rm Max}<u_0,q>=1$.
In no noise case, $z(t)=v(t)$ and
\be
o(t) =  <v,q> = \frac{M_v}{2\pi} \int_{-\infty}^{\infty} \frac{N_u |U_0(j \om)|^2}{S_t(\om)} \exp(j \om t)  d\om \,.
\ee
Thus, using the above normalization, we obtain ${\rm Max}[o(t)]= M_v$, from
which one can estimate the strength of the input GW $h_0$. The matched filter SNR can be evaluated as 
$
{\rm SNR^2} = ({\rm Max}[o(t)])^2/{\rm Var}(<n,q>)
$.
The variance of the filtered noise is given by
\be
{\rm Var}(<n,q>) = E\{<n,q>^2\} - (E \{<n,q>\})^2 = N_u \,.
\ee
Thus, the matched filter SNR is given by \footnote{For detailed discussion, see \cite{arch05}}
\be
\label{eq:snrM}
{\rm SNR_M^2} = \frac{M_v^2}{N_u} = \frac{1}{2 \pi} \int_{-\infty}^{\infty} \frac{|V(j\om)|^2}{S_t(\om)} d \om \,,
\ee
where $V(j\om)$ is the FT of $v(t)$. In terms of the signal and system poles $p_i, i=1,\ldots 6$ 
{\small
\be
\label{eq:p12}
\!\!\!\!\!\!\!\!\!\!\!\!\!\!\!\!\!\!\!\!\!\!\!
\!\!\!\!\!\!\!\!\!\!\!\!\!\!\!p_1 = -\om_+ + j/\t_+,~p_2 = -p_1^*,~p_3 = -\om_- + j/\t_-,~p_4 = -p_3^*,~p_5 = -\om_0 + j/\t,~p_6 = -p_5^* \, ,
\ee 
}

\noindent and the decay times $\t_+,\t_-$ pertaining to the response of the two modes [i.e. $f_\pm$, $\om_\pm = 2 \pi f_\pm$]
at the output of the delta filter \footnote{In \cite{Astone:1997dj}, $\t_+,\t_-$ are
indicated by $\t_{3+}, \t_{3-}$ respectively,}, the SNR$_{\rm M}$ is
\hspace{-1in}
{\small
\begin{displaymath}
\!\!\!\!\!\!\!\!\!\!\!\!\!\!\!\!\!\!\!\!\!\!\!\!\!\!\!\!\!\!\!\!\!\!\!\!
\!\!\!\!
{\rm SNR}_M^2= \frac{-h_0^2 L^2 \om_0^2 K^2}{16 S_n} 
\Re \left[\frac{\t_+ p_1^7}{\om_+ \prod_{i, i \neq 1,2} (p_1^2 - p_i^2)} 
+  \frac{\t_- p_3^7}{\om_- \prod_{i, i \neq 3,4} (p_3^2 - p_i^2)}
+ \frac{\t p_5^7}{\om_0 \prod_{i, i \neq 5,6} (p_5^2 - p_i^2)} \right] \,.
\end{displaymath}
}
\noindent

\section{Explorer-Nautilus Delta filter}
As stated earlier, the delta-filtering is the most natural approach
for detecting unknown short GW bursts in the narrow-band detector.
The delta filters are developed as follows.

The normalized system response to a delta-like signal (used to build
the delta-filter) is $u^\d(t)$ such that
$U^\d(j \om) = W_{ux}/M_\d$ \cite{Astone:1997dj} and ${\rm Max}[u^\d(t)] = 1$. 
Then, the delta filter, constructed from this response has the transfer
function 
%
$Q^\d(j\om) = N_\d U^{\d*}(j\om)/S_t(\om)$ 
%
with the filter normalization 
$N_\d = [\frac{1}{2\pi} \int_{-\infty}^{\infty} |U^{\d}(j\om)|^2/S_t(\om) d \om]^{-1}$.
This filter construction is such that if an impulse is 
incident on the bar with $v(t) = M_v u^\d(t)$ then the maximum of
the filtered output ${\rm Max}(<v,q^\d>)$ is $M_v$. 

However, when the response of the detector to the damped sinusoid
is filtered through the delta filter, the filtered output becomes
\bea
\label{eq:gdelta}
o(t) &=& <v,q^\d> = \frac{M_v}{2\pi} \int_{-\infty}^{\infty} \frac{N_\d U_0(j \om) U^{\d *} (j \om) }{S_t(\om)} \exp(j \om t)  d\om \,,\no \\
&=&  \frac{-h_0 L \om_0 K N_\d}{4\pi S_n M_\d} \int_{-\infty}^{\infty} \frac{\om^4 \exp(j \om t) d\om}{(\om^2 - p_1^2)(\om^2 - p_2^2)(\om^2 - p_3^2)(\om^2 - p_4^2)(\om - p_5) (\om - p_6)} \,. \no
\eea
We solve this integration by applying the residue theorem and obtain
\bea
\label{eq:gt2}
&o(t)& = \frac{-h_0 L \om_0 K N_\d}{8 S_n M_\d} \Re [-\frac{\t_+ e^{-t/\t_+}}{\om_+} \frac{p_1^5 e^{-j \om_+ t}}{\prod_{i=3,4} (p_1^2-p_i^2) \prod_{k=5,6} (p_1-p_k)} \no \\
&-&\frac{\t_- e^{-t/\t_-}}{\om_-} \frac{p_3^5 e^{-j \om_- t}}{\prod_{i=1,2} (p_3^2-p_i^2)\prod_{k=5,6} (p_3-p_k)} 
+ \frac{j e^{-t/\t}}{\om_0} \frac{p_5^6 e^{-j \om_0 t}}{\prod_{i=1,..,4} (p_5^2 - p_i^2)} ] \quad \,\,\, .
\eea

The variance of the filtered noise is ${\rm Var}(<n,q^\d>) = N_\d$. Thus,
\be
\label{eq:snrD}
{\rm SNR_\d^2} = \frac{M_v^2}{N_\d} ({\rm Max}<u_0,q^\d>)^2 \,.
\ee
 

\section{Comparison and Numerical plots}
To illustrate, we use the parameters pertaining to Explorer[Feb 2005];
two resonant frequencies $f_-=904.7$ Hz, $f_+=927.452$ Hz,
$\t_+ \sim 140$ ms, $\t_- \sim 150$ ms, $K \sim 1.66 \times 10^{13} {\rm V/m}$
and $S_n \sim 10^{-8} {\rm V^2/Hz}$.
\begin{figure}
\caption{\label{fig:exp} (a) $|F(j \omega)|~{\it vs}~f$ 
for $\tau = 50,250,450,650$ ms, the higher peak corresponds to
higher $\t$. (b) Explorer : Two sided PSD in per
$\sqrt{\rm Hz}$, (c) ${\rm SNR_M}$ {\it vs} $\tau$ for $h_0 \sim 10^{-20}$. 
}
\begin{tabular}{ccc}
\includegraphics[height=0.23\textheight,width=0.3333\columnwidth]{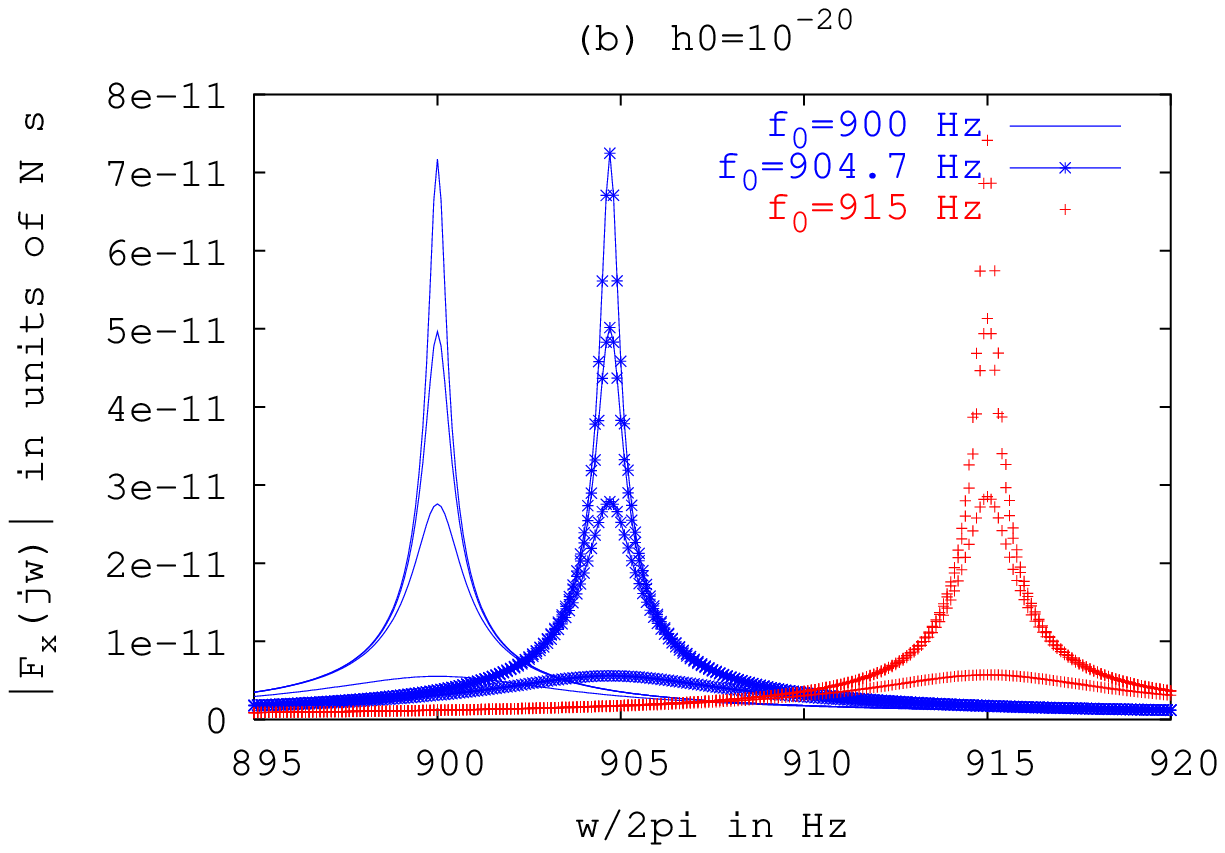}
\includegraphics[height=0.23\textheight,width=0.3333\columnwidth]{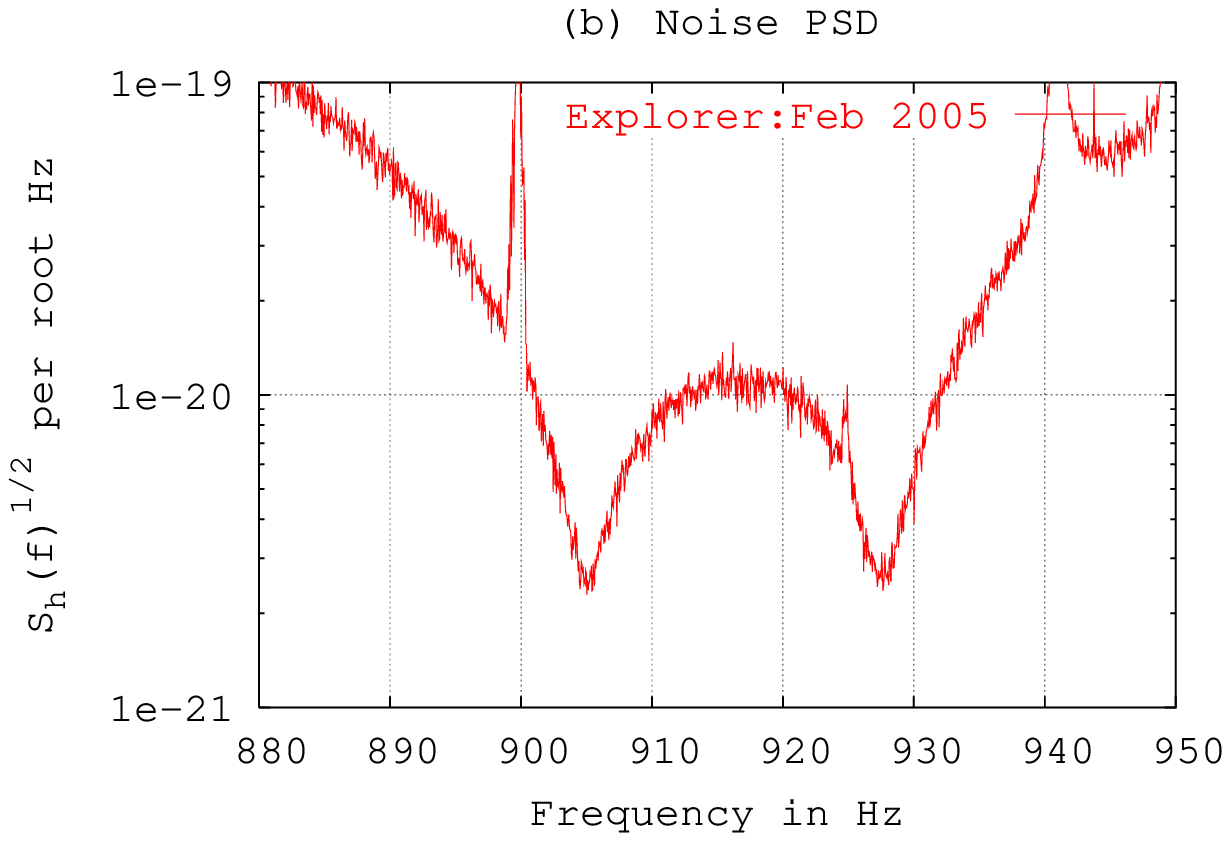}&
\includegraphics[height=0.23\textheight,width=0.3333\columnwidth]{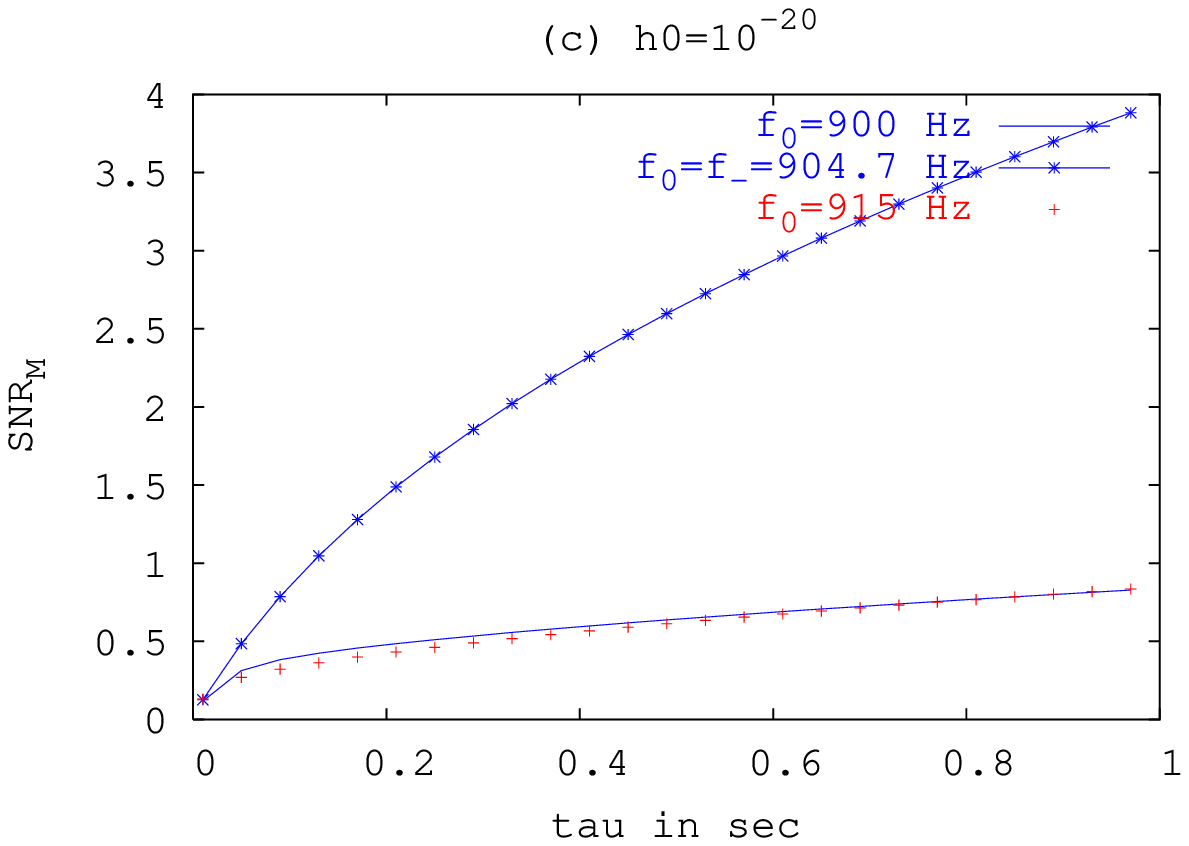}
\end{tabular}
\end{figure}
Fig.\ref{fig:exp}(b) shows the noise PSD of Explorer. The BW of the
detector at the level of $\sqrt{S_n} \sim 10^{-20}/\sqrt{Hz}$ is $\sim 20$ Hz
and we define the sensitive frequency band as ${\rm FB}:\{900, 932\}$.
We note that at
$f \sim 915$ Hz, the sensitivity is the worst within FB.
We divide our 
study in 2 cases, (a)$f_0 \in {\rm FB}$, (b)$f_0$ far from FB. 

Fig.\ref{fig:exp}(c) shows the plot of ${\rm SNR_M}$ {\it vs} $\tau$ for fixed 
$h_0 \sim 10^{-20}$. As $\tau$ increases ${\rm SNR_M}$ increases as the signal
spends more time in the detector. However, for $f_0$ close to $f_\pm$,
this increase is sharp as the incoming $h(t)$ excites the resonances 
and gives more and more energy to $f_\pm$ as $\t$ increases,
see Fig.\ref{fig:exp}(a).
For a given $\t$, the difference in SNR's of two incoming 
GW with frequencies $f_-$ and $915$ Hz is related to
the corresponding $S_h(f_-)$ and $S_h(915)$
[see Fig.\ref{fig:exp}(b)]. In Fig.\ref{fig:exp}(c), the SNR for
$f_0=900$ Hz and $f_0=915$ Hz are similar as the detector sensitivity 
is similar at those frequencies. For $f_0$ away from the resonance,
the detector band falls in the tail of the signal Lorentzian giving
small power to the resonances even at high $\t$. Thus, the increase in
SNR is very slow in such case.
The similar plot can be obtained fixing the signal energy instead of
$h_0$. However, the signal energy of QNM is itself a function of $\tau$.
Thus, for clear demonstration of dependence of SNR on $\tau$, we 
fix $h_0$ here.

\subsection{${\rm SNR}_M$ {\it vs} ${\rm SNR}_\d$}
\begin{figure}[htb]
\caption{\label{fig:snra} SNR loss in (a) Case (a), (b) Case (b), (c) Error in timing}
\begin{tabular}{ccc}
\includegraphics[height=0.23\textheight,width=0.33\columnwidth]{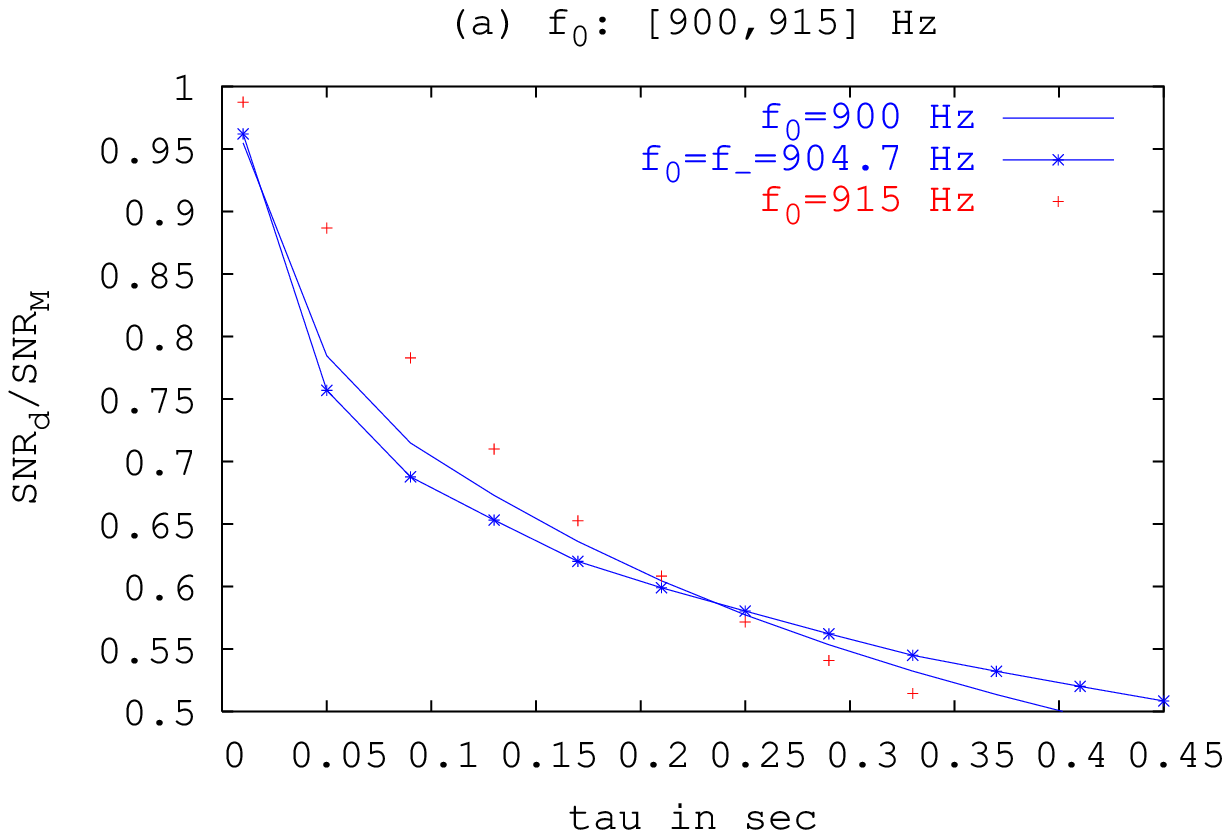}&
\includegraphics[height=0.23\textheight,width=0.33\columnwidth]{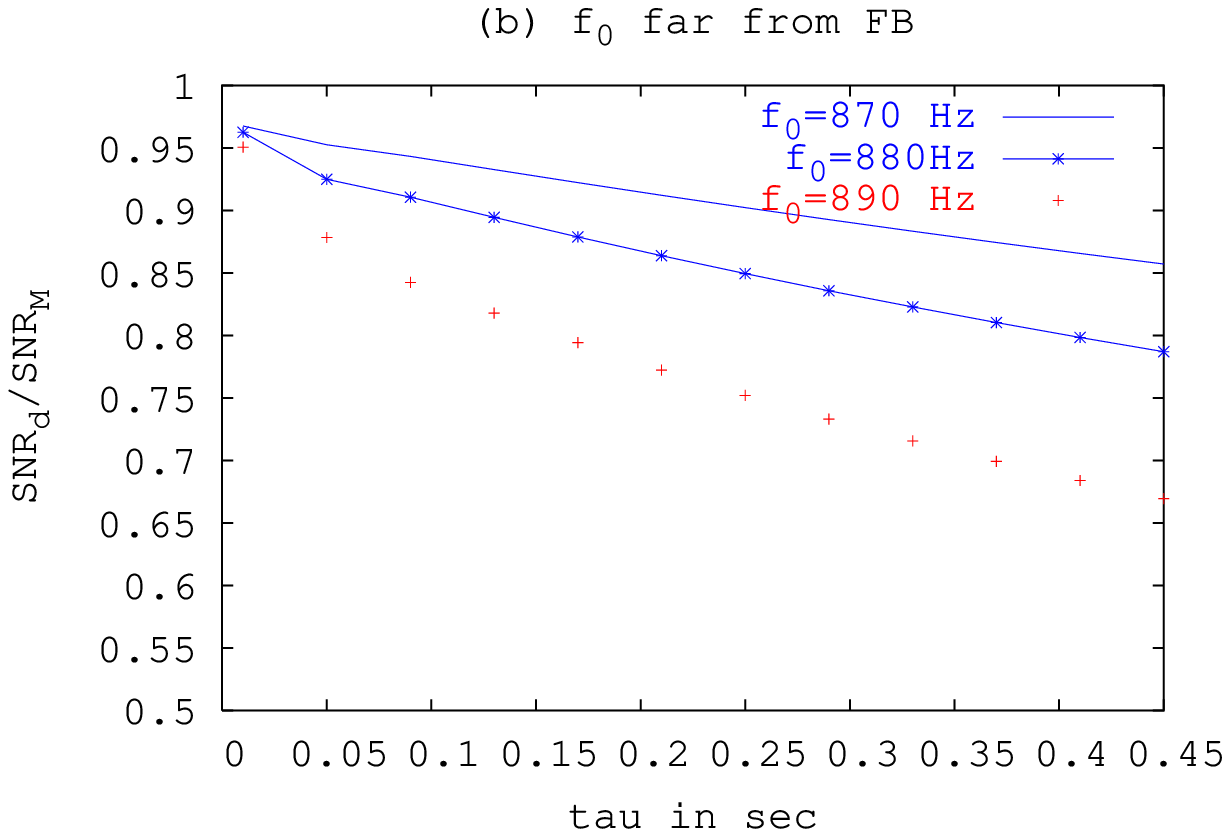}&
\includegraphics[height=0.23\textheight,width=0.33\columnwidth]{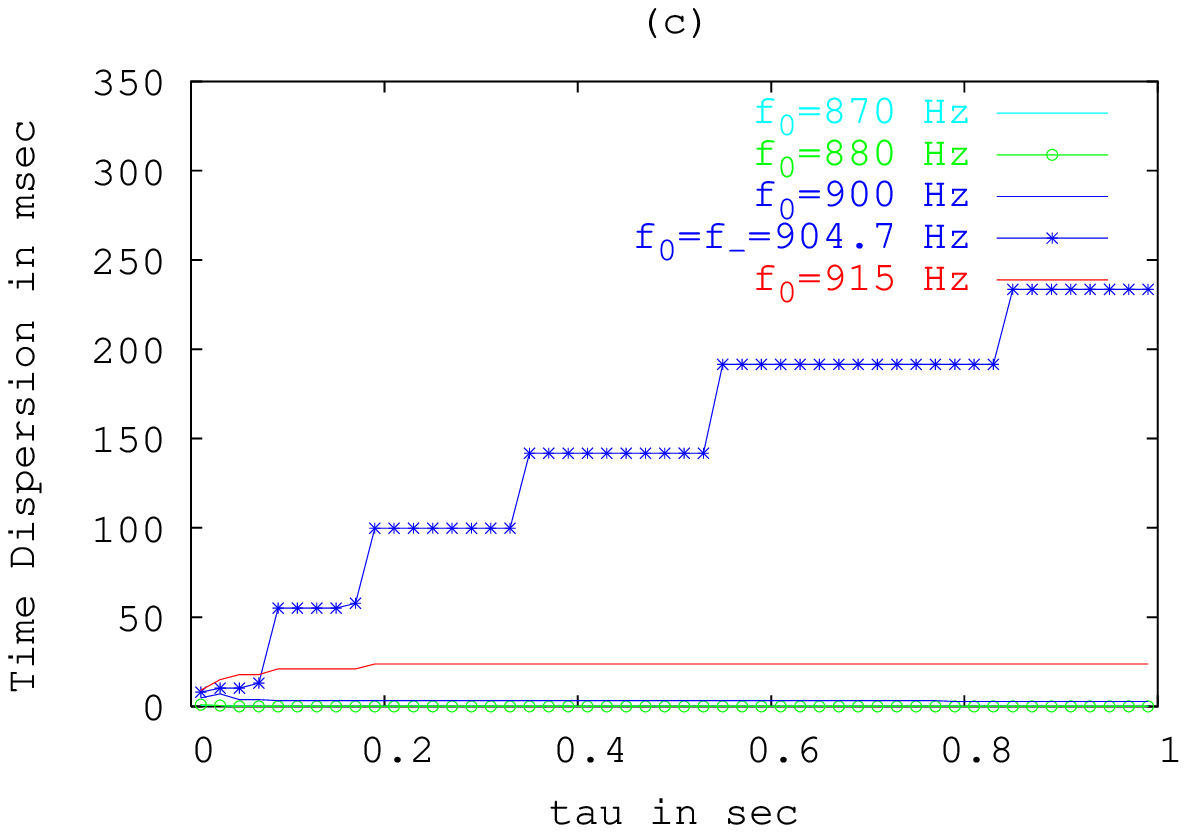}
\end{tabular}
\end{figure}

To validate the delta filter, we study the loss in SNR when the signal is filtered with the
delta filter, i.e. the ratio between Eq.(\ref{eq:snrD}) and Eq.(\ref{eq:snrM}).
In Fig. \ref{fig:snra}, we plot this ratio for case (a) and (b)
respectively. We show that when $f_0 \in FB$, the ${\rm SNR_\d}$ is comparable to that of the ${\rm SNR_M}$ (assuming the SNR loss of $\sim 15\%$) for all $\tau < 50$ ms. This loss increases as $\t$ increases. Thus, when $f_0 \in {\rm FB}$, for
high $\t$,
the delta filter is far from optimal.
Contrarily, when $f_0$ is far from FB, 
the tail of the signal Lorentzian gives relatively flat spectrum in detector
BW (for small values of $\t$), similar to a delta-like signal. Consequently, delta filter matches the 
signal
giving ${\rm SNR_M}$ comparable to that of ${\rm SNR_\d}$ for 
$\t$ as large as $\sim 200$ ms. However, as $\t$ increases, due to the
nature of signal Lorentzian, the energy given to both the resonances is not
the same. This results in the decrease in SNR ratio as $\t$ increases even 
for case (b). This loss is related to the error in the estimation of $h_0$. However, we note that 
in case(b), the ${\rm SNR_M}$ is well below the ${\rm SNR_M}$ obtained for
case(a) for the same $h_0$.

\subsection{Arrival timing error}
\begin{figure}[htb]
\caption{\label{fig:filt}Normalized output of the (A) matched filter, (B) delta filter, for $\tau=10 ms,0.2 s, 1 s$ and $f_{0}=f_{-}$.
}
\begin{tabular}{ccc}
\includegraphics[height=0.14\textheight,width=0.333\columnwidth]{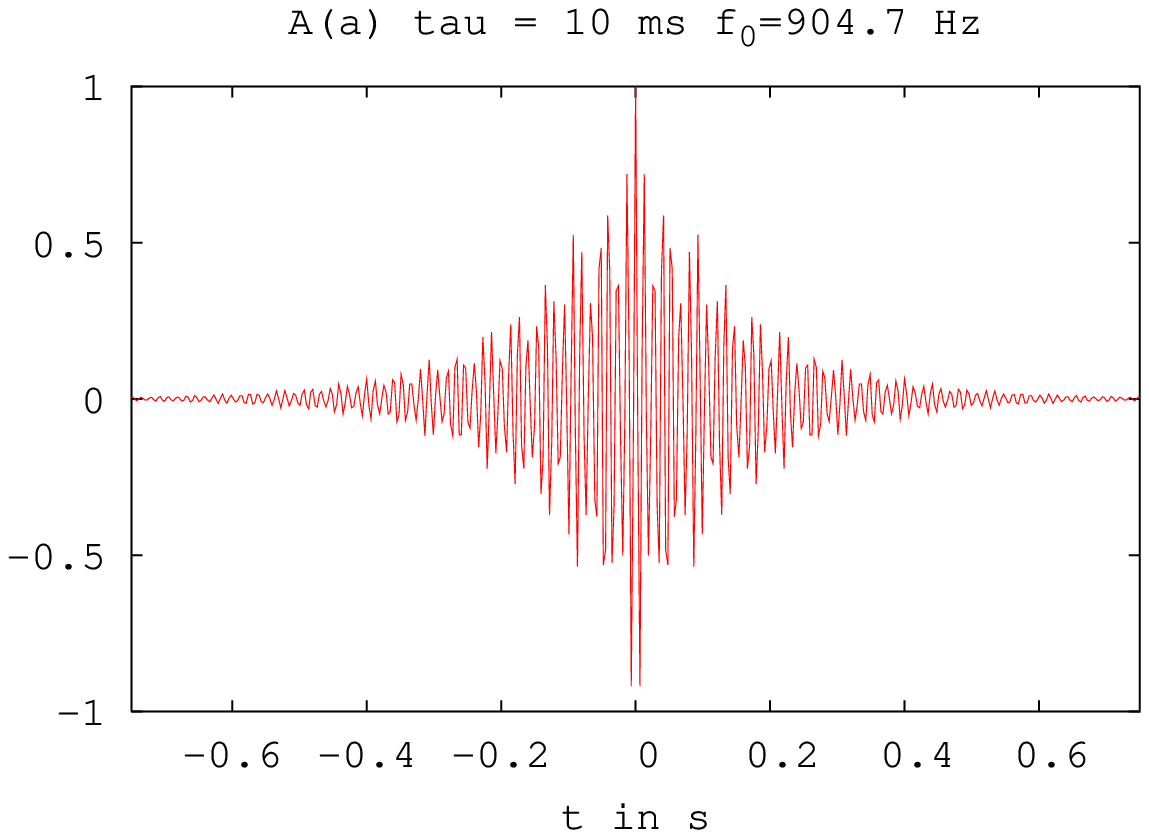}&
\includegraphics[height=0.14\textheight,width=0.333\columnwidth]{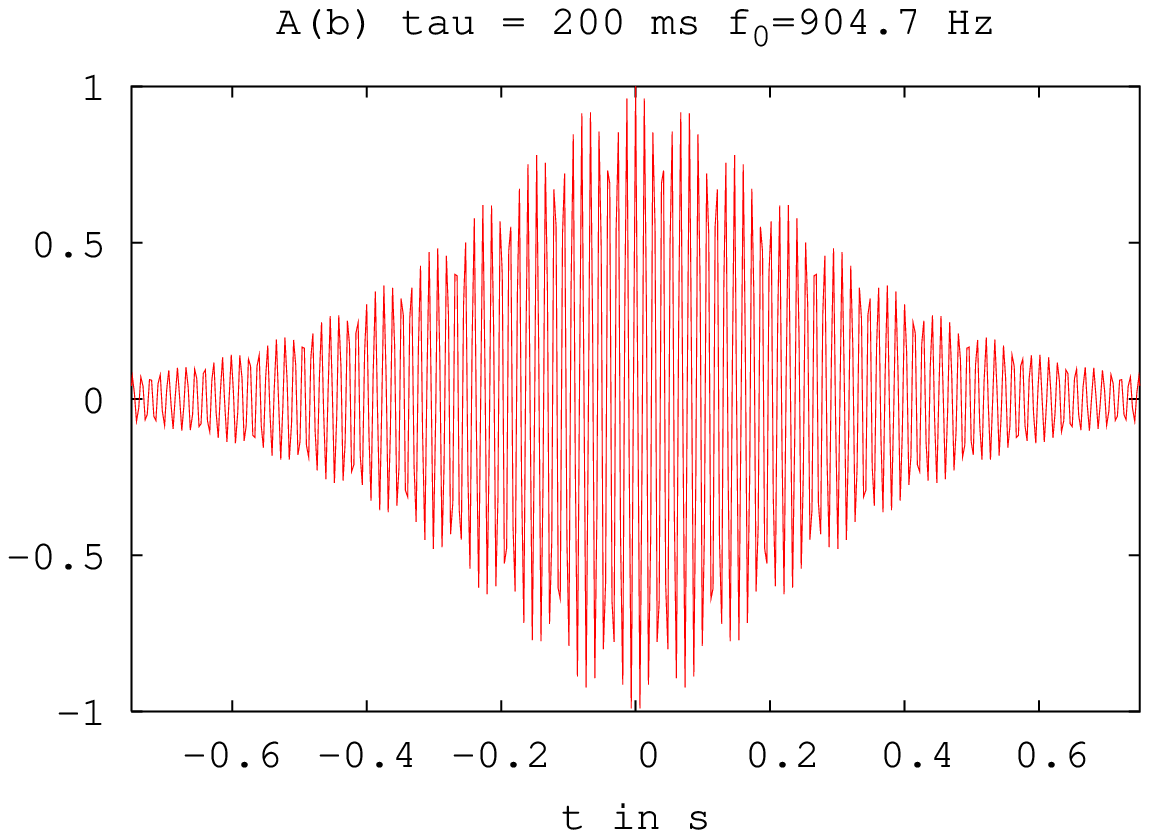}&
\includegraphics[height=0.14\textheight,width=0.333\columnwidth]{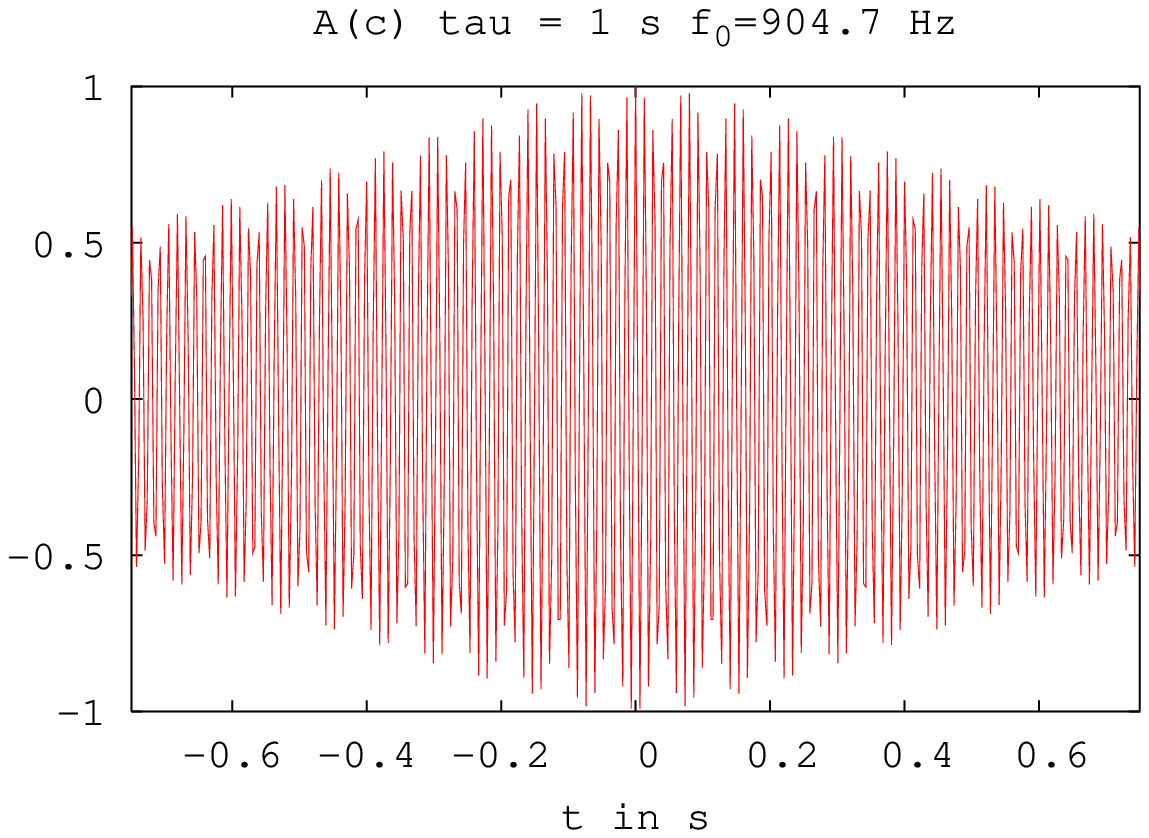}\\
\includegraphics[height=0.14\textheight,width=0.333\columnwidth]{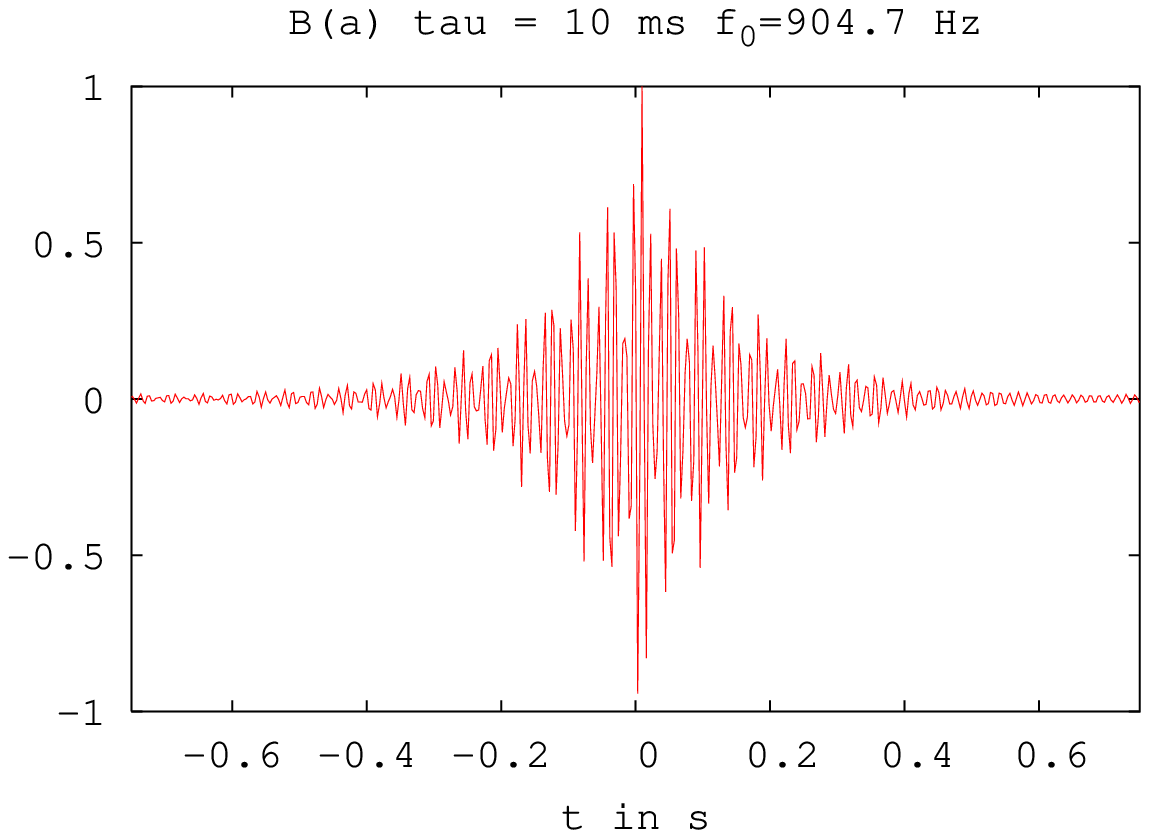}&
\includegraphics[height=0.14\textheight,width=0.333\columnwidth]{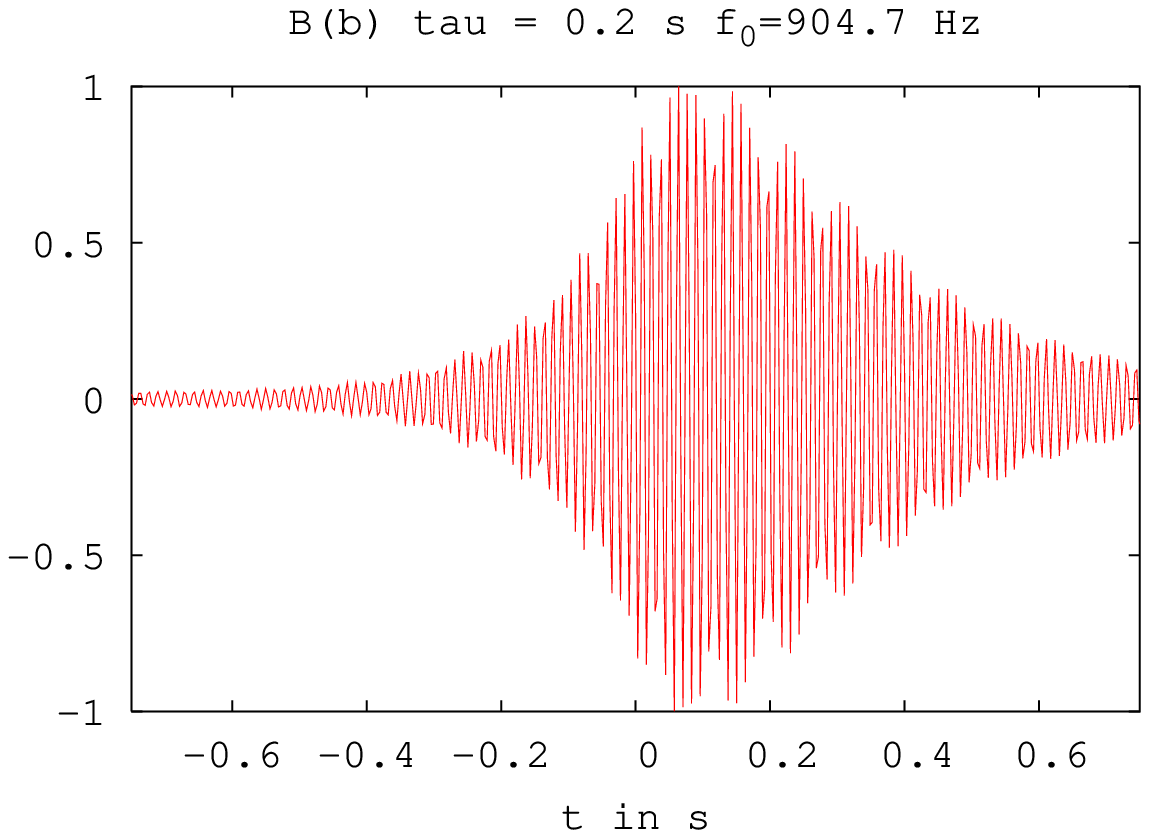}&
\includegraphics[height=0.14\textheight,width=0.333\columnwidth]{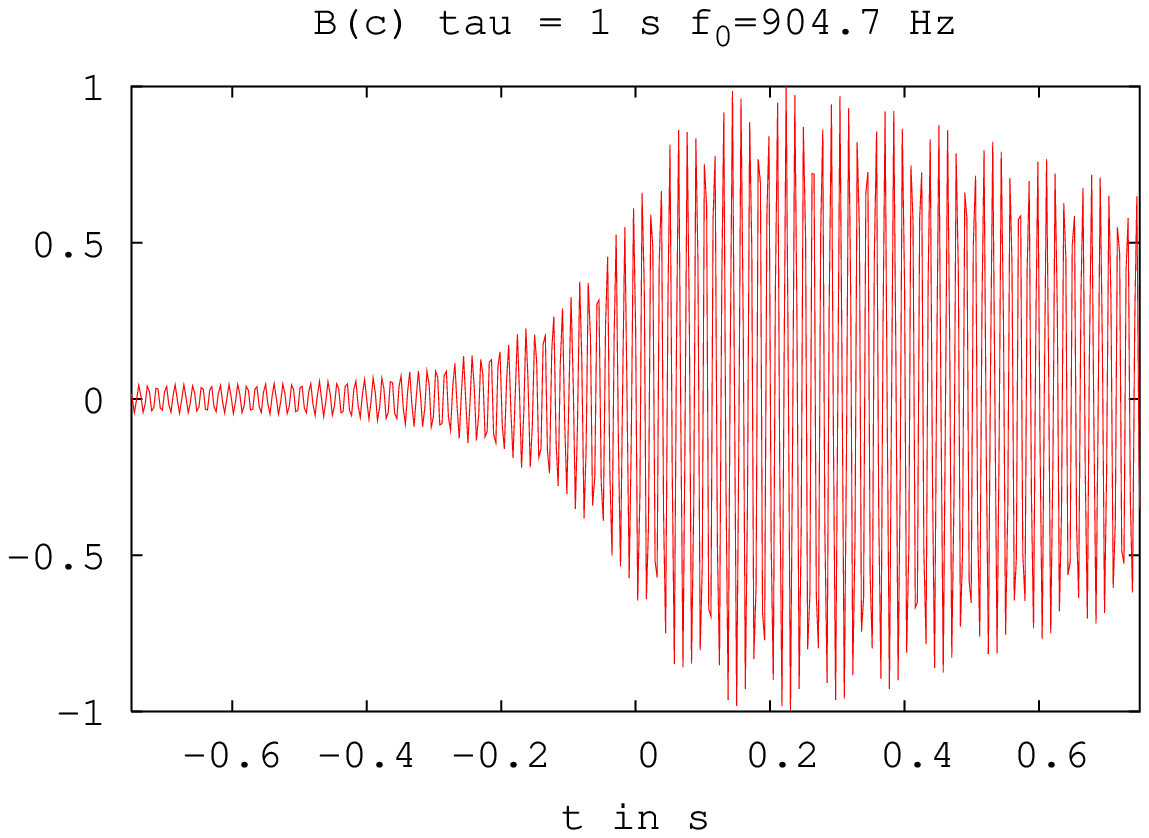}
\end{tabular}
\end{figure}

In Fig. \ref{fig:filt}, we plot the output of the delta as well as
matched filter for case (a) with $f_0 = f_-$ and $\t = 0.01,0.2,1$
s. We show in Fig. \ref{fig:filt} A (a), when $\t < \t_\pm$, the decay
of the matched filtered output is dominated by the $\t_\pm (\sim 140
{\rm ms})$.  However, as $\t$ increases and is $> \t_\pm$ [see
Fig \ref{fig:filt} A (b) and (c)], the decay time of the filtered
output is dominated by the signal term which can increase the arrival
timing error.  However, it is worth noting that with the noise, this
error not only depends on the decay time of the filtered output but
also on the SNR. As $\t$ increases, the decay time increases but at
the same time the ${\rm SNR_M}$ also increases (which we wish to
investigate in future with simulations).

The output of the delta filter is shown in Fig.\ref{fig:filt} B. The
output is asymmetric about $t=t_0=0$ as the delta filter is causal [as
it is not properly matched]
unlike the matched filter. Mathematically, it can be seen by the
the relative sign difference between the signal term
and the two resonances in Eq.(\ref{eq:gt2}). In this case,
the filtered output becomes maximum when $t \sim t_0+\t$. Thus,
the arrival timing error is proportional to $\t$, see fig. \ref{fig:snra} (c).
The steps correspond to the beating frequency which in general depends on
$f_0$ and $f_\pm$. In this case when $f_0 = f_-$, it is
$(f_+-f_-)/2 \sim 12$ Hz. We note that this beating is
crucial while fixing the coincidence
timing window while performing coincidences
between say Explorer-Nautilus or Explorer-Virgo.

In case (b), when $f_0$ is far away from FB, the situation is contrary.
In this case, the timing error is small, [see Fig. \ref{fig:snra} (c)].
As explained earlier, for $f_0$ away from FB and low $\t$, the signal acts as a delta-like 
signal. As a result, delta-filter itself is a matched filter hence gives
no timing error. Mathematically, the signal term in Eq.(\ref{eq:gt2}) is small compared to other terms. 

\section{Conclusion}
In this work, we did a comparative study of the response of the matched filter
{\it vs} delta-filters for filtering the damped sinusoids GW signals 
for resonant bar detectors. We divided our study in two cases:
(a) $f_0 \in {\rm FB}$ and (b) $f_0$ far from the FB with 
${\rm FB} =\{900, 932\}$ using Explorer configuration. 
We find that in case (a), the loss in SNR increases as $\t$ and so does
the arrival timing error if
delta-filter is used instead of matched filter. However, in case (b), the
signal almost acts like a delta for small $\t$ hence the SNR loss is negligible
for small $\tau$ however as $\t$ increases, the SNR loss gradually increases.
The arrival timing error is minimal for all $\t$. Thus, we can optimally
use delta filters for detecting signals in case (b) for $\tau$ as large as
$200$ ms.
But, to detect damped sinusoids in case (a) with $\t$ as large as $50$ ms,
it is mandatory to use the 
optimal filtering as opposed to delta-filter. This poses a problem of setting
an ``optimal'' grid of templates in $f_0, \t$ when $f_0, \t$ are fixed.
However, as described earlier, $f_0$ and $\t$ can evolve 
in the detector bandwidth. For detecting such signals and perform 
coincidences, the delta filter is inadequate and an optimal ``matched''
filter is difficult to construct due to insufficient knowledge of signal
waveform. Alternative detection methods are needed which we pursue
in future work.

\ack AP would like to thank INFN for financial support, and ICTP, LNF for the hospitality.

\end{document}